\documentclass{iau}

\usepackage{amsmath}
\usepackage{graphicx}
\usepackage{multirow}
\usepackage{hyperref}

\newcommand{\ca}{\mbox{Ca\,{\sc ii}~K\,}}
\newcommand{\ccmax}{CC$_{\rm max}$}

\newcommand{\aap}{    {\it Astron. Astrophys.}}

\newcommand{\apj}{    {\it Astrophys. J.}}

\newcommand{\mnras}{  {\it Mon. Not. Roy. Astron. Soc.}}

\newcommand{\solphys}{{\it Solar Phys.}}

\begin{document}

\lefttitle{Dibya et al.}
\righttitle{Solar Rotation Profile of the Chromosphere}

\jnlPage{1}{7}
\jnlDoiYr{2023}
\doival{10.1017/xxxxx}

\aopheadtitle{Proceedings IAU Symposium}
\editors{Alexander Getling \&  Leonid Kitchatinov, eds.}

\title{Differential Rotation of the Solar Chromosphere using multidecadal \ca\ Spectroheliograms}
\author{Dibya Kirti Mishra$^{1,2}$, Srinjana Routh$^{1,2}$, Bibhuti Kumar Jha$^{3}$, Subhamoy Chatterjee$^3$ and Dipankar Banerjee$^{1,4,5}$}
\affiliation{$^1$Aryabhatta Research Institute of Observational Sciences, Nainital-263002, Uttarakhand, India\\ email: \email{dibyakirtimishra@aries.res.in}}
\affiliation{$^2$Mahatma Jyotiba Phule Rohilkhand University, Bareilly-243006, Uttar Pradesh, India}
\affiliation{$^3$Southwest Research Institute, Boulder, CO 80302, USA}
\affiliation{$^4$Indian Institute of Astrophysics, Koramangala, Bangalore 560034, India}
\affiliation{$^5$Center of Excellence in Space Sciences India, IISER Kolkata, Mohanpur 741246, West Bengal, India \\ email: \email{dipu@aries.res.in}}

\begin{abstract} 
The study of the differential rotation in the chromosphere of the Sun is of significant importance as it provides valuable insights into the rotational behaviour of the solar atmosphere at higher altitudes and the coupling mechanism between the various layers of the solar atmosphere. In this work, we employed the image correlation technique, explicitly focusing on plages, intending to estimate the chromospheric differential rotation. For this purpose, we have utilized \ca\ spectroheliograms (1907\,--\,2007) from the Kodaikanal Solar Observatory (KoSO), recently calibrated with a better technique to ensure accuracy. Our analysis indicates that plages in the chromosphere exhibit faster rotation and a smaller latitudinal gradient when compared to the rotation rate obtained through sunspot tracking. Furthermore, we investigate the temporal analysis of the chromospheric differential rotation parameters across various solar cycles. 
\end{abstract}

\begin{keywords}
Sun: chromosphere, Sun: plages, Sun: solar cycle, Sun: differential rotation 
\end{keywords}

\maketitle

\section{Introduction}\label{sec:intro}

After the discovery of sunspots by Galileo and other astronomers, the concept of solar differential rotation first came into existence in 1630 by Christoph Scheiner. Since then, it has been the subject of interest in the scientific community and the extensive study to describe this with the help of sunspot observations led to the development of the well-established empirical relation
\begin{equation}
\Omega= A + B\sin^2{\theta} + C\sin^4{\theta},
\label{eq1}
\end{equation}
where $\Omega$ represents the angular velocity, $\theta$ denotes the heliographic latitude, while $A$ represents the equatorial rotation rate, and $B$ and $C$ correspond to latitudinal gradients. Gradually, advancements in technology have enabled extensive investigations into the solar differential rotation, utilizing various methods such as sunspot tracking \citep{jha2021, Jha2022thesis}, spectroscopy technique\citep{Howard1970}, and helioseismology \citep{Komm2008}. Despite the numerous studies conducted in both the photosphere and the interior of the Sun, our understanding of differential rotation behaviour in the higher atmospheric layers, including the chromosphere and corona, remains limited.

As previously mentioned, sunspots are employed as tracers to explore photospheric differential rotation owing to their prolonged persistence. Similarly, plages are also considered promising candidates for investigating differential rotation phenomena in the chromosphere due to their prolonged existence \citep{Bertello2020}. These extended chromospheric features are predominantly observed in \ca\ spectral line centered at $393.367$\,nm. However, it is important to note that plages differ from sunspots in that they undergo relatively rapid morphological changes, making it challenging to employ traditional tracking methods on them. Despite this limitation, there have been various attempts to analyze the rotation profile in the chromosphere by tracking plages over relatively short time spans \citep{Antonucci1979}. In addition to these, more recent efforts have employed advanced techniques such as Fast Fourier Transform \citep{Singh1985}, image cross-correlation \citep{Bertello2020}, and autocorrelation method \citep{li2023} to gain insights into this phenomenon.

Despite the previous works, our comprehension of chromospheric differential rotation and its connection to the photosphere remains incomplete. The Kodaikanal Solar Observatory (KoSO), which possesses an extensive archive of \ca\ spectroheliograms (1904\,--\,2007), presents a distinctive opportunity for investigating chromospheric differential rotation at various phases of the solar cycle. In this article, we will discuss the image correlation method employed for finding the chromospheric rotation profile by utilizing newly calibrated \ca\ data of KoSO.

\section{Data and Methodology}
The archival \ca\ dataset over a century (1904--2007) collected by KoSO\footnote{The digitized data are available at \url{https://kso.iiap.res.in/data}.} provides extensive information about the chromospheric activity. These data were acquired on photographic plates/films using the diffraction grating system of the bandpass of 0.05\,nm centered at 393.367\,nm \citep{Chatterjee2016, Jha2022thesis}. To enhance their quality, the dataset has undergone digitization and advanced calibration procedures were applied to ensure accuracy, as described in \citet{Chatzistergos2020} with details. Furthermore, \citet{Jha2022thesis} devised a reliable approach to enhance the precise alignment of \ca\ images spanning from 1907 to 2007, a critical factor for the accurate determination of differential rotation. For our purpose of estimating differential rotation in the chromosphere, we employed the recently calibrated and rotation-corrected \ca\ data (1907\,--\,2007) of KoSO.

To account for the dynamic nature of plages, as elaborated in Section~\ref{sec:intro}, we have implemented the image correlation technique to determine chromospheric differential rotation. In this process, we select a pair of observations while imposing a constraint on the time interval $\Delta t$ having condition: 0.5\,days $< \Delta t <$ 1.5\,days. Subsequently, these images are transformed into the heliographic grid \citep{thompson2006} with a resolution of $0.1^\circ$ per pixel. Next, we partition the data into 5$^\circ$ latitudinal bands, as exemplified by A1 and A2 in Fig.~\ref{method}(a), within the range of $\pm55^\circ$ in both latitude and longitude. This approach mitigates the impact of projection effects near the solar limb, ensuring more accurate and robust measurements.

Following this, we employ a 2D image correlation technique on selected latitude bands, where we specify a range for the shift in both latitude [$\Delta\theta_0\pm 1^\circ$] and longitude [$\Delta\phi_0\pm 2^\circ$]. In this context, $\Delta\theta_0$ is set to 0, and $\Delta\phi_0$ is determined as $\Omega(\theta)\Delta t$ based on the work by \citet{jha2021}. This process yields optimal values for the longitudinal shift, denoted as $\Delta\phi$, which corresponds to the highest Correlation Coefficient (\ccmax) and is depicted in the brown curve of Fig.~\ref{method}(a). These values are instrumental in computing $\Omega_{\rm synodic}$, defined by the equation
\begin{equation}
    \Omega_{\rm synodic} = \frac{\Delta\phi}{\Delta t}.
    \label{eq2}
\end{equation}
In the process of computing $\Omega_{\mathrm{synodic}}$, we apply a conditional criterion for which we consider values of $\Delta\phi$ where CC exceeds 0.2. In cases where CC falls below this threshold and there is no discernible variation in CC, such values are excluded from the analysis. After the computation of $\Omega_{\mathrm{synodic}}$ as per Eq.~\ref{eq2}, we perform a sidereal correction to account for the Earth's motion around the Sun. This correction is accomplished using the equation followed in \citet{jha2021} for synodic to sidereal conversion. This procedure enables us to calculate $\Omega$ ($\Omega{\mathrm{sidereal}}$) successfully for each latitudinal band, encompassing the entire data period spanning from 1907 to 2007.

\section{Result}

\begin{figure}[htb!]
  \centering
  \includegraphics[scale=0.46]{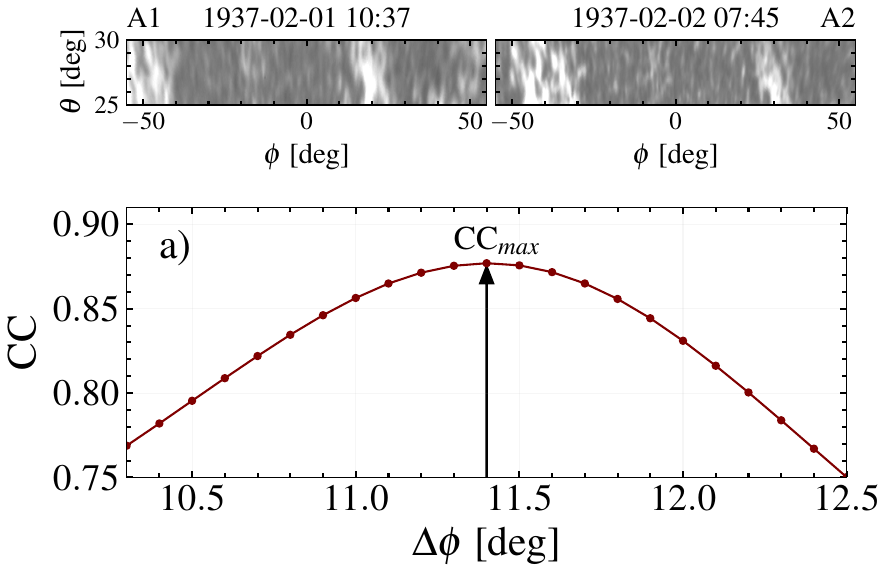}
  \label{method}
  \includegraphics[scale=0.6]{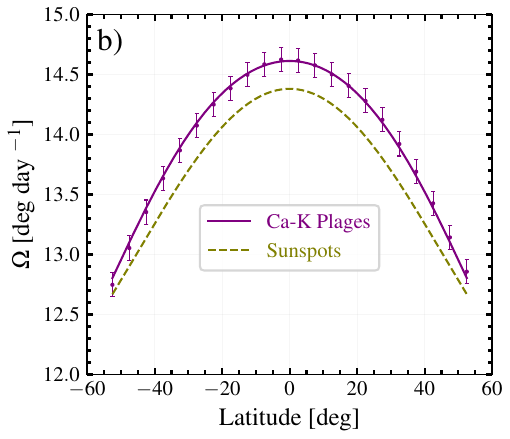}
  \label{main_result}
  \caption{a) The latitude bands for two consecutive days observed on 1937-02-01 10:37 IST (A1) and 1937-02-02 07:45 IST (A2), within the latitude range of $+25^\circ$ to $+30^\circ$. The brown curve in (a) illustrates the variation in CC with $\Delta\phi$ for a constant $\Delta\theta$, a product of a 2D cross-correlation technique applied on A1 and A2, showing the location of the maximum Cross-Correlation value (\ccmax). b) The comparative representation of chromospheric rotation using plages (purple continuous curve) and sunspot rotation profile (olive dashed line) obtained from \citet{jha2021}.}
  
\end{figure}

To determine the chromospheric differential rotation, we calculate the angular velocity $\Omega$ for 22 latitude bands ranging from -55$^{\circ}$ to +55$^{\circ}$ by taking the mean value over the entire data period (purple dots; Fig.~\ref{main_result}(b)). The associated errors in these $\Omega$ values encompass both statistical errors ($\sigma_{{\rm SSE}}$) and least count errors ($\sigma_{\rm LCE}$). Notably, $\sigma_{\rm LCE}$ due to the resolution of the heliographic grid ($0.1/\Delta t$) predominantly contributes to the total error. To quantify the differential rotation within the chromosphere, we employed a fitting procedure involving Eq.~\ref{eq1} applied to the average angular velocity ($\Omega$). This analysis yielded the following parameter values: $A$ = 14.61\,deg day$^{-1}$, $B$ = -2.18\,deg day$^{-1}$, $C$ = -1.10\,deg day$^{-1}$ with corresponding uncertainty of $\pm$0.04, $\pm$0.37, $\pm$0.61 respectively \citep{mishra2023}. We then compare the rotation rate of plages (purple continuous curve) with the sunspot rotation rate \citep[olive dashed line;][]{jha2021} obtained using KoSO White-Light data spanning from 1923 to 2011, as can be seen in Fig.~\ref{main_result}(b). The comparison reveals that the rotation rate of plages is greater than that of the sunspots, thus indicating faster rotation of the chromosphere using plage as a tracer compared to the photosphere based on sunspot tracking.

\begin{figure}[htb!]
  \centering
  \includegraphics[scale=0.58]{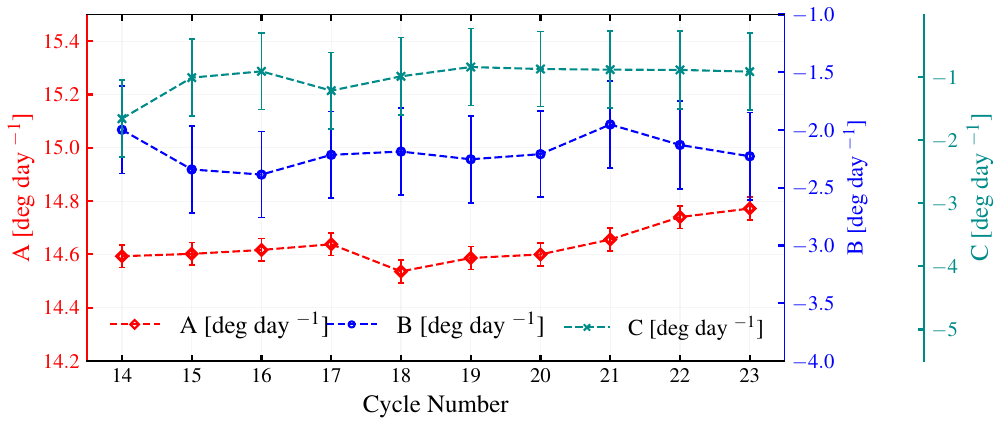}
  \caption{The variation in differential rotation parameters (A, B, and C) throughout solar cycles.}
  \label{solar_cycle}
\end{figure}

We also explore the temporal evolution of these differential rotation parameters over the course of solar cycles, as shown in Fig.~\ref{solar_cycle}. The analysis reveals that the chromospheric differential rotation remains insignificantly variant across solar cycles. Notably, when examining the variation of parameter A (red) concerning solar cycle number (Fig.~\ref{solar_cycle}), we observe an increasing trend in A values following solar cycle number 20. This could be attributed to a degradation in data quality during the latter half of the dataset, as discussed in previous studies \citep{mishra2023}.

\section{Conclusion}

In summary, our analysis indicates that plages exhibit a higher rotation rate compared to sunspots, characterized by a less latitudinal gradient. Furthermore, it suggests that this differential rotation of plages does not change significantly throughout solar cycles. These findings align with recent studies by \citet{li2023}, which also support the observation of a faster rotation in the chromosphere than the photosphere. In addition to these, there are also a few studies done recently in the higher solar atmosphere \citep[e.g.][]{sharma2020}, which also indicate faster rotation in the higher solar atmosphere. However, the underlying reasons for this higher rotation rate in the chromosphere remain not fully understood. In the future, we intend to undertake a more comprehensive investigation into solar rotation across various heights within the solar atmosphere to deepen our understanding of this phenomenon. 

\section{Acknowledgment}
We thank IAU for the travel grant awarded to us, making it possible to attend the symposium in person in Armenia.



\end{document}